\documentstyle[12pt,amssym,aaspp4,flushrt]{article}
\def\Msolar{{M$_{\odot}$\,}}

\received{Jun 10, 1999} 
\accepted{ }

\slugcomment{To appear in ``The Astrophysical Journal''}

\begin{document}

\title{On the Globular Cluster IMF below 1\,\Msolar\altaffilmark{1}}

\altaffiltext{1}{Based on observations with the NASA/ESA {\it Hubble
Space Telescope}, obtained at the Space Telescope Science Institute,
which is operated by AURA, Inc., under NASA contract NAS5-26555}

\author{Francesco Paresce and Guido De Marchi} 
\affil{European Southern Observatory, Karl-Schwarzschild Strasse 2, 
D--85748 Garching, Germany\\
fparesce@eso.org, demarchi@eso.org}
\authoremail{fparesce@eso.org, demarchi@eso.org}

\begin{abstract}
Accurate luminosity functions (LF) for a dozen globular clusters have
now been measured at or just beyond their half-light radius using HST.
They span almost the entire cluster main sequence (MS) below ~
$0.75$\,\Msolar. All these clusters exhibit LF that rise continuously
from an absolute $I$ magnitude $M_I\simeq 6$ to a peak at $M_I \simeq
8.5-9$ and then drop with increasing $M_I$.  Transformation of the LF
into mass functions (MF) by means of the mass luminosity (ML) relations
of Baraffe et al. (1997) and Cassisi et al. (1999) that are consistent
with all presently available data on the physical properties of low
mass, low metallicity stars shows that all the LF observed so far can
be obtained from MF having the shape of a log-normal distribution with
characteristic mass $m_c=0.33 \pm 0.03$\,\Msolar and standard deviation
$\sigma = 1.81 \pm 0.19$. In particular, the LF of the four clusters in
the sample  that extend well beyond the peak luminosity down to close
to the Hydrogen burning limit (NGC\,6341, NGC\,6397, NGC\,6752, and
NGC\,6809) can only be reproduced by such distributions and not by a
single power-law in the $0.1 - 0.6$\,\Msolar range. After correction
for the effects of mass segregation, the variation of the ratio of the
number of higher to lower mass stars with cluster mass or any simple
orbital parameter or the expected time to disruption recently computed
for these clusters by Gnedin \& Ostriker (1997) and Dinescu et al.
(1999) shows no statistically significant trend over a range of this
last parameter of more than a factor of $\sim 100$. We conclude that
the global MF of these clusters have not been measurably modified by
evaporation and tidal interactions with the Galaxy and, thus, should
reflect the initial distribution of stellar masses. Since the
log-normal function that we find is also very similar to the one
obtained independently for much younger clusters and to the form
expected theoretically, the implication seems to be unavoidable that it
represents the true stellar IMF for this type of stars in this mass
range.
\end{abstract}
 

\section{Introduction}

The IMF is a critical ingredient in our understanding of a large number
of basic astronomical phenomena such as the formation of the first
stars, galaxy formation and evolution, and the determination of the
absolute star formation rate.  It also plays a dominant role in any
star formation theory as the end result of molecular cloud contraction
and fragmentation. Moreover, the IMF is one of the important factors
determining the rate of cluster disruption via internal and external
evolution (relaxation and tidal shocking) and, in consequence, of the
possible dark matter content of galaxy halos. In this latter context, a
single power law IMF increasing as $dN/dm \propto m^{-\alpha}$ with
$\alpha \gtrsim 2$ all the way to very low substellar masses is
required to substantially affect the baryonic mass budget of the halo
(Chabrier \& M\'era 1997; Graff \& Freese 1996).

The actual measurement of a MF is a complex process whose ultimate
precision and reliability rests heavily on a very careful quantitative
analysis of all sources of possible random and systematic error.  The
basic uncertainties presently stem mainly from sample contamination,
incompleteness, errors in the mass-luminosity and color-magnitude
relations, the age, distance, and extinction of the stars, their
evolution, mass segregation, and unresolved binaries. The IMF itself
depends crucially, in most cases, on knowledge of the age and on any
subsequent effects such as external dynamical evolution of a cluster in
a galactic tidal field. For all these reasons, it has proven very
difficult to pin down the shape of the IMF observationally with the
required reliability and accuracy in a wide variety of stellar
environments (Scalo 1998, 1999). The slope of the MF at the lowest mass
end of the stellar MS and, in particular, whether or not there is a
turn-over at the lowest masses before the H-burning limit and whether
or not the IMF is universal or rather depends on the initial physical
conditions in the natal environment are critical open issues at
present.

Globular clusters represent, in principle, the ideal sample from which
to deduce the stellar IMF and properly answer the above questions. They
offer a large statistically significant sample of relatively bright,
coeval, equidistant stars with, in most cases, relatively small
variations of chemical composition and extinction within each cluster.
They were all formed very early in the history of the Galaxy and there
is no evidence of subsequent star formation episodes.  The binary
fraction outside the core is less than $10 - 15\,\%$ and has an
insignificant effect on the measured LF (Rubenstein \& Bailyn 1999).
Mass segregation is a relatively straightforward and well understood
phenomenon quantifiable by simple Michie--King models such as those
used by Meylan (1987, 1988). The only potentially serious obstacle is
related to the possible modification of the IMF by the effect of tidal
interactions with the Galaxy potential.  This interaction, integrated
over the orbit and time, is expected to slowly decrease the slope of
the global mass function of the cluster (Vesperini 1998) thereby
effectively masking the original IMF from our present day observations,
no matter how precise and detailed they are.

Since deep LF of a dozen globular clusters (GC) in our Galaxy have now
been accurately measured, we are in a good position to address
observationally the issue of if and, possibly, how the interaction
history of these clusters, whose Galactic orbits are reasonably well
known, affects their LF in the mass range where the signature is
expected to be most significant. In this paper, we show that LF
obtained at or just beyond the half-light radius of these clusters
surveyed are completely insensitive to this history and that they can
indeed be used to deduce an uncontaminated stellar IMF below
$1$\,\Msolar for these stars.

\section{Observational Data}

The main characteristics of the data used for this study are summarized
in Table\,1 and their relevant presently available physical parameters
are listed in Table\,2. All the listed clusters have well measured LF
in the critical range $6.5<M_I<10$ and some even well beyond these
limits.  We have restricted this study to absolute $I$ magnitudes
greater than $4.5$ corresponding to a mass of $\sim 0.75$\,\Msolar to
avoid the mass range near the turn-off where cluster age and instrument
saturation might significantly affect the determination of the LF
(Silvestri et al. 1998; Baraffe et al. 1997; De Marchi et al. 1999).
The LF of these clusters in number per $0.5$ magnitude bins as
determined by analysis of their color--magnitude diagrams (CMD) are
plotted with the relevant $1\,\sigma$ error-bar in Figure\,1, as a
function of the absolute $I$-band magnitude obtained using the distance
moduli given in Table\,2.

The statistical errors shown in Figure\,1 have been determined by
combining in quadrature the uncertainty resulting from the Poisson
statistics of the counting process with that accompanying the
measurement of the photometric incompleteness. The family of curves
shown in Figure\,1 represents a very homogeneous sample of objects all
observed and analyzed with the same basic techniques well outside the
core in a region where the low-mass MS is well populated. The most
obvious feature of the observed LF is the peak located at $M_I\simeq
8.5-9$ with a rising and descending part on each side.  Only the
clusters NGC\,6341, NGC 6397, NGC\,6656, NGC\, 6752, and NGC\,6809
extend significantly beyond $M_I=10$ in this sample due to the
difficulty of obtaining reliable LF at such faint luminosities for the
more distant objects.

\begin{table} 
\caption{The clusters in our sample. Columns 3 and 4 show the radial
distance at which the LF have been measured, respectively in arcmin
and in units of the half-light radius ($r_h$; see Table\,2). LF have
been measured within $\pm 1\,r_h$ of the given average position $r$.}
\begin{tabular}{rlrrl}
\hline\hline
NGC & Name & \multicolumn{1}{c}{$r$} & \multicolumn{1}{c}{$r/r_h$} &  Reference\\
\hline
 104 & 47\,Tuc       & 4.6 &  1.6 &  De Marchi \& Paresce (1995b)\\
5139 & $\omega$\,Cen & 4.6 &  0.9 &  De Marchi (1999)\\
5272 & M\,3          & 1.5 &  1.5 &  Marconi et al. (1997), Carretta et al. (1999)    \\
6121 & M\,4          & 6.1 &  1.3 &  Pulone, De Marchi \& Paresce (1999)\\
6254 & M\,10         & 2.4 &  1.3 &  De Marchi \& Paresce (1996)   \\
6341 & M\,92         & 4.6 &  4.5 &  Piotto, Cool \& King (1997)   \\
6397 &               & 4.6 &  1.8 &  Paresce, De Marchi \& Romaniello (1995)\\
6656 & M\,22          & 2.6 &  0.8 &  De Marchi \& Paresce (1997) \\
6752 &                & 3.1 &  1.5 &  Ferraro et al. (1997)        \\
6809 & M\,55          & 2.5 &  0.9 &  De Marchi \& Paresce (1996)   \\
7078 & M\,15          & 4.6 &  4.6 &  De Marchi \& Paresce (1995a)  \\
7099 & M\,30          & 4.6 &  4.6 &  Piotto, Cool \& King (1997)   \\
\hline
\end{tabular}
\end{table}

\begin{deluxetable}{rccccrrccrrccc}
\tablewidth{43pc}
\tablecaption{Clusters' structural parameters}
\tablehead{
\colhead{NGC} & \colhead{(m-M)$_I$} & \colhead{$r_h$} & \colhead{$r_c$} &  
\colhead{$c$} &  \colhead{$Z_G$} & \colhead{$R_G$} & \colhead{$P$} &
\colhead{[Fe/H]} & \colhead{$T_d$} & \colhead{$T_d$} & \colhead{$m_c$} & \colhead{$\sigma$} & \colhead{$\Delta\,\log\,N$}\\
(1)&(2)&(3)&(4)&(5)&(6)&(7)&(8)&(9)&(10)&(11)&(12)&(13)&(14)\\}
 
\startdata
 104 & 13.37 & 2.9 & 0.2 & 2.48 &  3.2 &  7.4 & 5.3 & -0.71 & 88 & 131 & 0.34 & 2.11 & 0.131 \nl 
5139 & 13.68 & 4.9 & 2.6 & 1.15 &  1.3 &  6.3 & 1.2 & -1.59 & 40 &  16 & 0.37 & 1.96 & 0.122 \nl  
5272 & 15.03 & 1.2 & 0.4 & 1.89 & 10.0 & 12.0 & 5.5 & -1.66 & 213& 275 & 0.36 & 1.61 & 0.251 \nl  
6121 & 12.11 & 4.5 & 1.2 & 1.53 &  0.5 &  6.2 & 0.7 & -1.33 & 13 &   2 & 0.32 & 1.84 & 0.208 \nl 
6254 & 13.59 & 1.9 & 0.7 & 1.66 &  1.7 &  4.7 & 3.4 & -1.60 & 22 &  23 & 0.33 & 2.19 & 0.136 \nl 
6341 & 14.41 & 1.0 & 0.3 & 1.65 &  4.3 &  9.1 & 1.4 & -2.24 & 30 &  33 & 0.30 & 1.73 & 0.104 \nl  
6397 & 12.00 & 2.9 & 0.1 & 1.69 &  0.5 &  6.0 & 3.1 & -1.91 &  4 &   4 & 0.32 & 1.73 & 0.237 \nl
6656 & 12.93 & 3.3 & 1.2 & 1.70 &  0.4 &  5.1 & 2.9 & -1.75 & 31 &  29 & 0.33 & 1.73 & 0.218 \nl
6752 & 13.18 & 2.0 & 0.5 & 2.15 &  1.8 &  5.1 & 4.8 & -1.54 & 35 &  96 & 0.42 & 1.84 & 0.089 \nl 
6809 & 13.57 & 2.7 & 1.7 & 1.27 &  1.9 &  4.2 & 1.8 & -1.82 & 14 &  11 & 0.32 & 1.68 & 0.250 \nl  
7078 & 15.19 & 1.0 & 0.1 & 1.77 &  4.8 & 10.5 & 5.5 & -2.17 & 48 & 155 & 0.30 & 1.50 & 0.140 \nl 
7099 & 14.43 & 1.0 & 0.1 & 2.40 &  5.4 &  6.8 & 3.2 & -2.13 & 23 &  40 & 0.30 & 1.84 & 0.159 \nl

\enddata
\tablecomments{Columns are as follows: (1) NGC number; (2) distance
modulus in the $I$ band defined as $(m-M)_V + 0.48 A_V$, with the
latter two values taken from Djorgovski (1993); (3) half-light radius
in arcmin (Djorgovski 1993); (4) and (5) core radius in arcmin and
concentration ratio (Webbink 1985); (6) and (7) distance in kpc
respectively from the Galactic plane and center (Djorgovski 1993); (8)
perigalactic distance in kpc (Dinescu et al. 1999); (9) metallicity
(Djorgovski 1993); (10) and (11) time to disruption in Gyr (assuming
$T_0=10$\,Gyr) respectively from Gnedin \& Ostriker (1997) and Dinescu
et al. (1999); (12) and (13) average characteristic mass (solar units)
and standard deviation of the log-normal distribution that best fits
the MF; (14) logarithmic ratio $\Delta\,\log\,N$ of lower to higher
mass stars as defined in Section\,3.}

\end{deluxetable}

\begin{figure}[h]
\plotfiddle{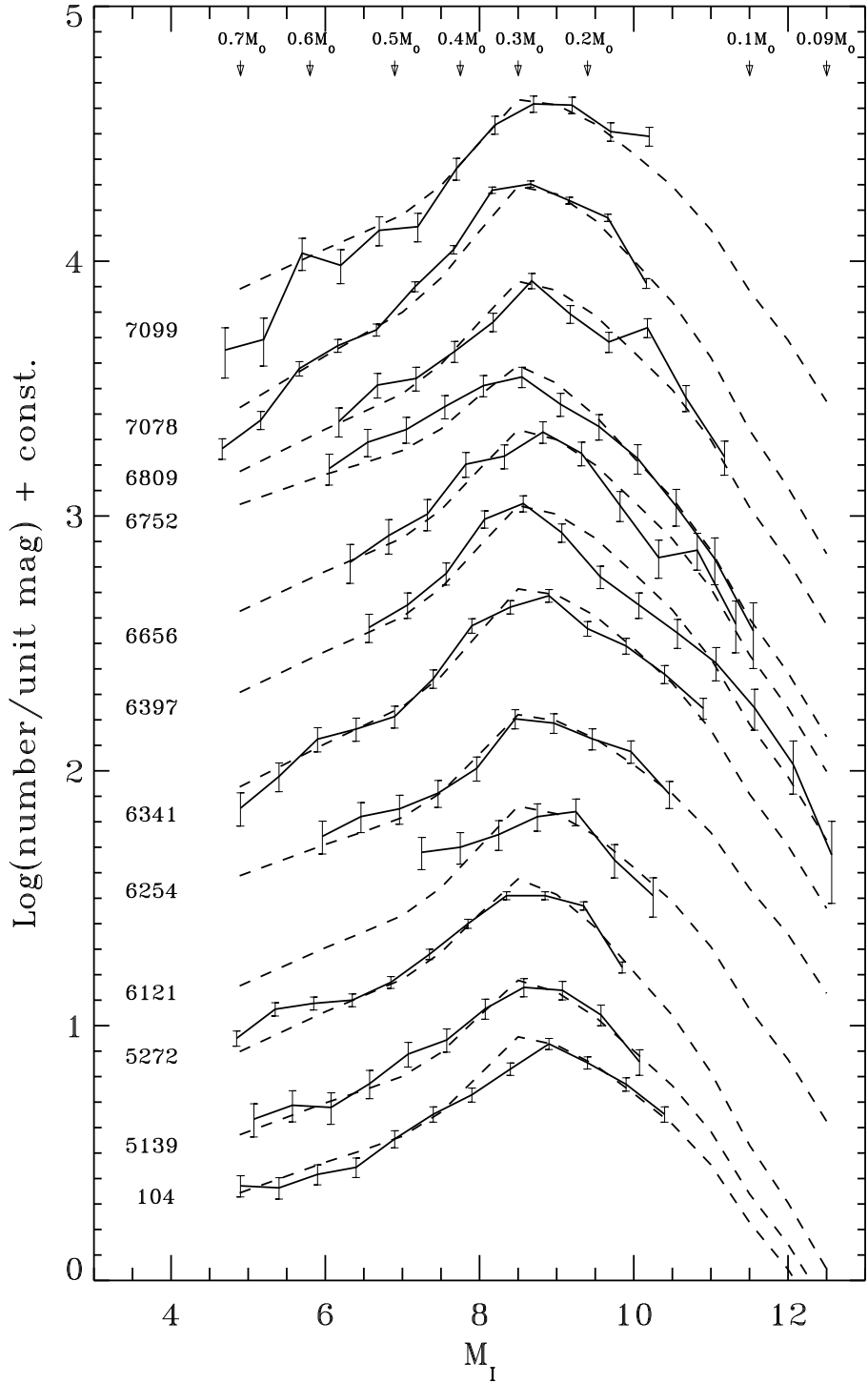}{20cm}{0}{80}{80}{-250}{0} 
\caption{Luminosity functions of the clusters in our sample (see
Table\,1). The data have been shifted vertically by an arbitrary
amount for enhanced visibility. The mass values shown at the top are
taken from Baraffe et al. (1997) for $[M/H]=-1.5$. The dashed lines
show the result of folding the log-normal MF of Figure\,3 through the 
ML relation of Baraffe et al. (1997; see Section\,3)} 
\end{figure}

\section{Conversion to a Mass Function}

The observed local LF (i.e. $dN/dM_I$) shown in Figure\,1 can be
converted into the corresponding MF (i.e. $dN/d\log m$) by the
application of a mass-luminosity relation (ML) as follows:

\begin{equation}
dN/dM_I = dN/d\log m \times d\log m/dM_I 
\end{equation}
    
Thus, the observed LF is simply the product of the MF with the
derivative of the ML relation. The critical step here, therefore, is
intimately connected to the proper realization of the appropriate ML
relation for the age and metallicity of the cluster. A number of
possibilities exists presently but the most reliable are the
theoretical ML relations explicitly computed for the appropriate
observational bandpasses by Alexander et al.  (1997), Silvestri et al.
(1998), Baraffe et al. (1997, 1998) and Cassisi (1999). Subtle
differences between the calculations can be considerably amplified by
the derivative process that is required to transform a LF into a MF and
vice versa as shown in Equation\,1. The main reason for these
differences lies in the use of the gray atmosphere approximation in the
first two models while the Baraffe et al. (1997) and Cassisi 
(1999) approach relies on a self-consistent non-gray model atmosphere
to provide the correct boundary conditions for the interior integration
(Chabrier, Baraffe, \& Plez 1996). Another reason is probably connected
to the differing equations of state used by the different authors. In
any case, the very fact that there are significant differences in the
various approaches that could definitely affect the final
transformation strongly argues that we should use the models that adopt
the fewest approximations to the physical processes underlying the
emitted spectrum and that adequately fit the widest possible range of
available data on low mass, low metallicity stars with the minimum of
adjustable parameters.  Presently, this advantage lies with the Baraffe
et al. (1997; see Chabrier et al. 1999 for the most recent review) and
Cassisi (1999) models that we, therefore, will use exclusively
in the following discussion. The two models are, fortunately,
practically indistinguishable from one another in the I band and our
mass range.  This consistency between independent determinations
increases our confidence in the basic reliability of the M-L relation
used here.

Since it is common practice to derive a MF from a given LF by dividing
the latter by the derivative of the ML relation, in principle we could
apply Equation\,1 to the data in Figure\,1 and derive the MF directly
through the inversion of the LF. In this way, however, the contribution
of the experimental errors and of the uncertainties inherent in
the models would become difficult to disentangle in the final result.
Instead, we prefer to assess the validity of a model MF by converting
it into the observational plane and comparing directly the resulting LF
with the observed one, precisely as indicated by the formalism of
Equation\,1.  Particular care should be used, however, when assuming a
functional form for the MF. Since the MF is often defined as the
differential probability distribution of stellar mass $m$ per unit
logarithmic mass range, i.e.  $F(\log m)$, it is convenient to use the
logarithmic index $-x = d \log F(\log m) / d \log m$ to characterize the
MF slope locally, over a narrow mass interval. As Scalo (1998) points
out, however, this characterization might seem to imply that the MF is
a power law with fixed index $x$, and thus it might seem to justify
extending over a wide mass range an index derived over a much narrower
mass interval.  This assumption is probably responsible for much of the
confusion still affecting the shape of the MF of GC stars near the
H-burning limit.
 
\begin{figure}[h]
\plotfiddle{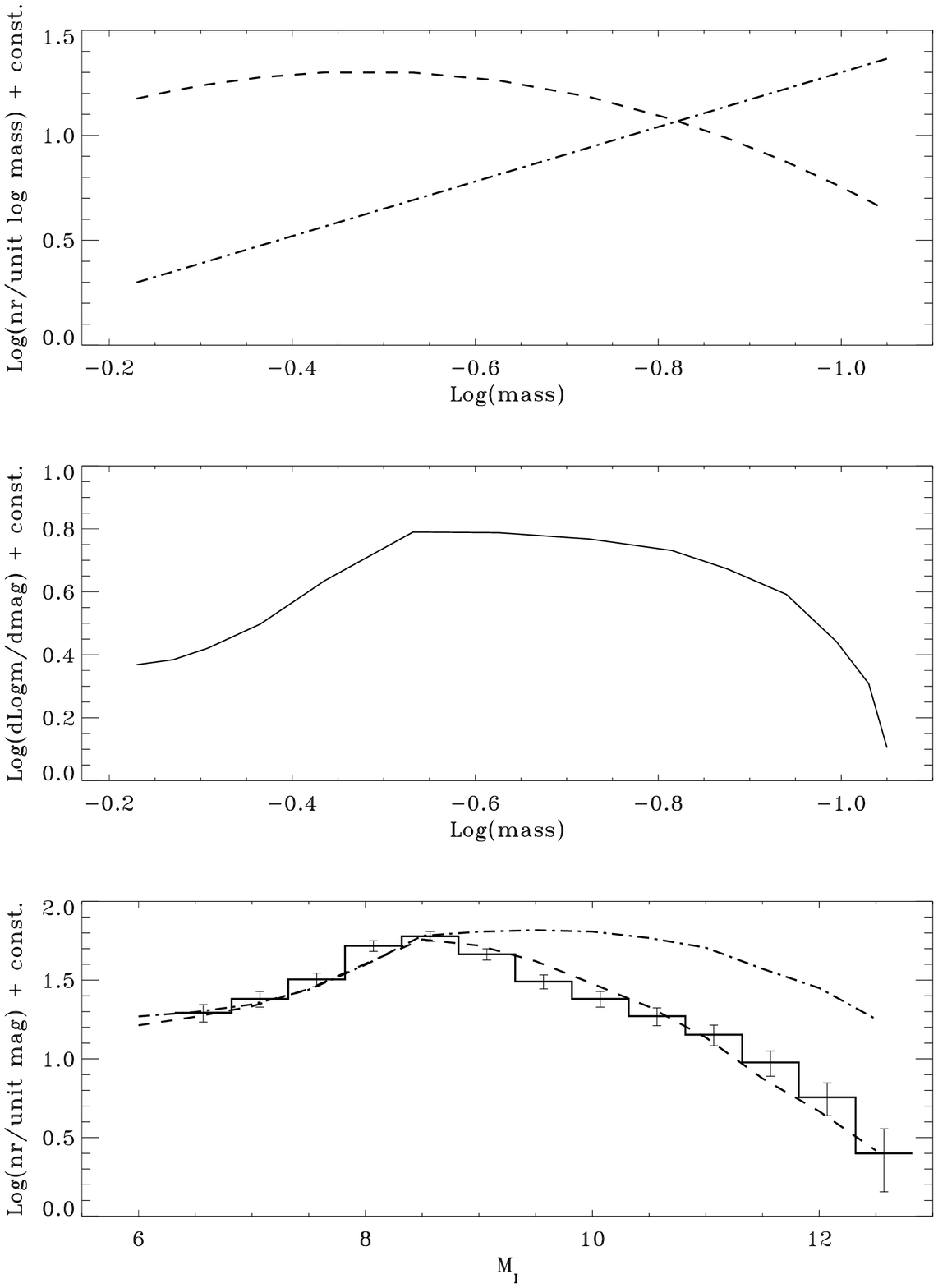}{17cm}{0}{80}{80}{-250}{-50}
\caption{Conversion of a mass function into a luminosity function
through the mass-luminosity relation. {\it Top panel:} two MF are used,
namely a power-law distribution (dot-dashed line, $x=1.3$) and a
log-normal distribution with $m_c = 0.32$ and $\sigma = 1.73$ (dashed
line). {\it Middle panel:} derivative of the M-L relation of Baraffe et
al. (1997) for $[M/H]=-1.5$.  {\it Bottom panel:} Once multiplied by
the derivative of the M-L relation only the log-normal MF (dashed line)
fits both the rising and falling portions of the luminosity function
simultaneously, whereas the power-law form (dot-dashed line) can only
fit one of them, depending on the choice of the exponent.}
\end{figure}

In fact, as recently shown by De Marchi, Paresce, \& Pulone (2000), the
deepest LF available for NGC\,6397 rules out the possibility that its
MF is represented by a power-law distribution with a single value of
the exponent $x$. We refer the reader to that paper for a complete 
discussion of the details, but report here the main points of that
derivation for sake of clarity:

\begin{enumerate}

\item 
The expected V and I magnitudes and colors from Baraffe et al. (1997,
1998) appropriate to the distance, age, and metallicity of NGC\,6397
fit very well the observed optical and near-IR CMD with no adjustable
parameters;

\item
The LF of this cluster has now been observed by three different
techniques from the optical to the near IR by two different groups all
of which give the same result throughout the wide luminosity range
between $6.0 \lesssim M_I \lesssim 12.5$ within observational errors;

\item
The resulting MF cannot be reproduced by a single power-law function
over this range as shown in Figure\,2, because after multiplication
with the appropriate ML relation such a function cannot simultaneously
fit both the rising and descending portions of the LF;

\item
The mass function that best fits the combined data on NGC\,6397 is a
log-normal distribution (see Equation\,2) with mass scale $m_c \simeq
0.3$ and standard deviation $\sigma \simeq 1.7$. The LF to which this
MF gives rise is shown as a dashed curve in the bottom panel of
Figure\,2.

\end{enumerate}

We cannot claim, of course, that all the other clusters in our study,
and especially those whose observed LF do not extend as far beyond $M_I
\simeq 10$ in Figure\,1 as NGC\,6397, have to have their MF in the shape
of that of NGC\,6397 necessarily, since we do not yet have data in the
fainter regime. They might have continuously rising MF down to the
H-burning limit as espoused by Chabrier \& M\'era (1997), for example,
but the ascending parts are essentially indistinguishable from one
another and the implication at least is that they would have a similar
behavior below the peak. This is certainly true for NGC\,6656 as
discussed in De Marchi \& Paresce (1997) as well as for NGC\,6341,
NGC\,6752, and NGC\,6809 since their LF extend to $M_I=11$ well beyond
the peak and are not compatible with an underlying power-law MF,
regardless of the value of its index $x$.

\begin{figure}[h]
\plotfiddle{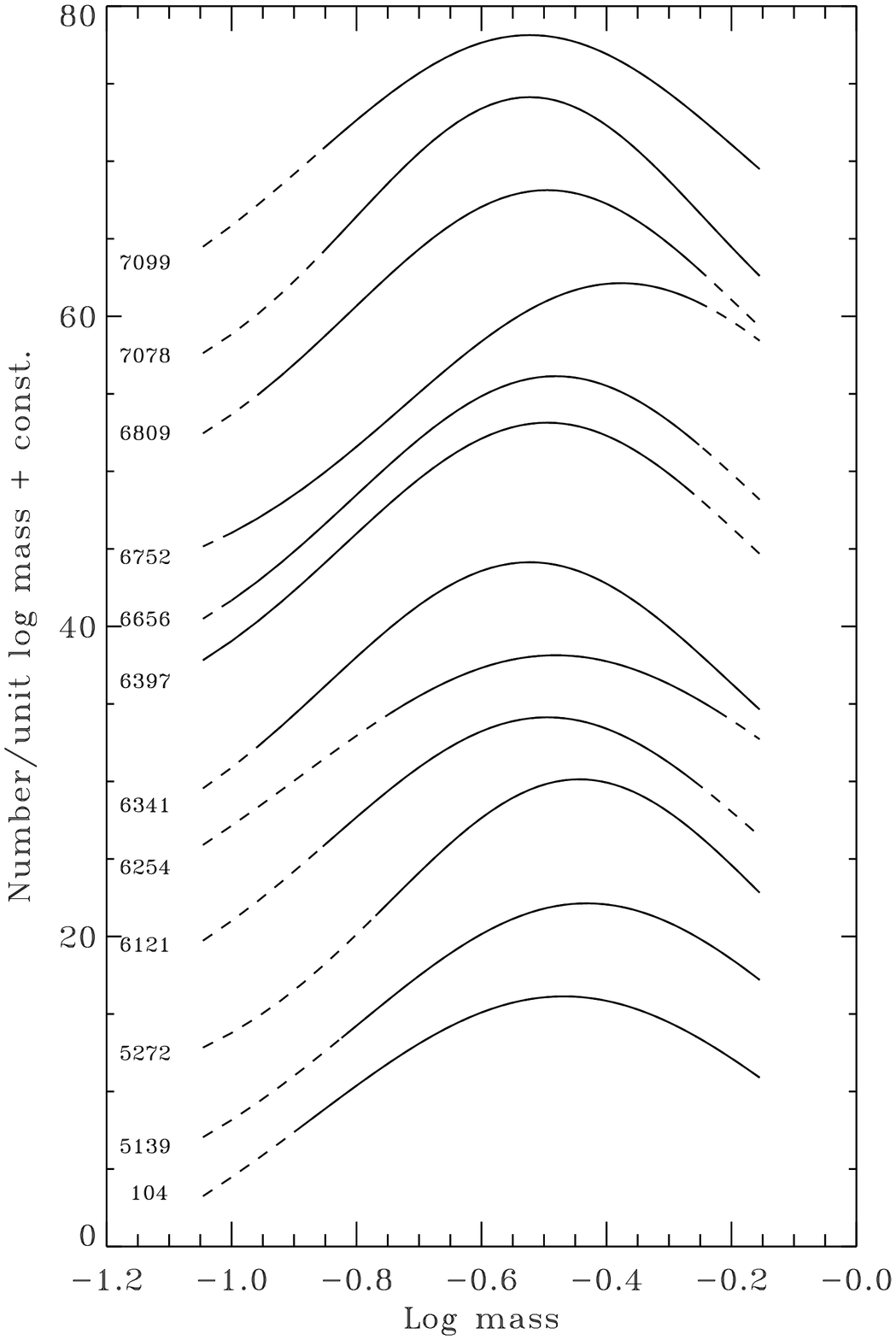}{20cm}{0}{80}{80}{-250}{0} 
\caption{Log-normal mass functions that best fit the LF shown in
Figure\,1. Note that the scale is linear along the y axis.}
\end{figure}

In Figure\,3, we show the log-normal distributions that accurately
reproduce the LF plotted in Figure\,1 over the whole magnitude range
spanned by the observations. Solid lines mark the portion of the MF
that have been fitted to the data, while the dashed lines represent the
extrapolation of the same MF to fill in the range $0.09 -
0.7$\,\Msolar. The dashed lines in Figure\,1 represent the theoretical
LF obtained with the MF shown in Figure\,3 that best fit the observed
LF.  The log-normal MF is characterized by only two parameters namely
the characteristic mass $m_{\rm c}$ and the standard deviation $\sigma$
and takes on the form:

\begin{equation}
\ln f \left( \log m \right) = A - \frac{\left[ \log(m/m_c)\right]^2}
{2 \sigma^2 }
\end{equation}

where $A$ is a normalization constant. The average values of the
parameters for this sample of clusters are $<m_c> = 0.33 \pm 0.03$ and
$<\sigma> = 1.81 \pm 0.19$. The uncertainties accompanying $<m_c>$ and
$<\sigma>$ represent the scatter of the individual values of $m_c$ and
$\sigma$ which are given for each cluster in Table\,2.  It should be
noted that the relatively small values of $\sigma$ in Table\,2 imply
that for $m < m_c$ the MF drops not only in the logarithmic plane, but
also in linear units, i.e. the number of stars per unit mass {\it
decreases with decreasing mass below the peak}. A simple, unbiased
measure of the steepness of the rise to the maximum of the MF shown in
Figure\,3 that does not depend on any preconceived notion on the shape
of the MF is $\Delta\,\log\,N$, defined as the logarithmic ratio of the
number of lower to higher mass stars taken from the MF in the mass
range between $m=m_c$ and $m=0.7$\,\Msolar. This is probably the most
convenient parameter to describe the region of the mass distribution
most likely to be affected by external and internal dynamics and is
listed in Table\,2 for each cluster. Another advantage of $\Delta\,\log\,N$
is that it is defined in a mass range where the stellar surface
structure is best understood and all presently available models for the
ML relation are in good agreement with each other (Silvestri et al.
1998) and is, in consequence, least likely to be subject to
uncertainties due to the LF to MF conversion.
 
\section{Correction for Mass Segregation}

\begin{figure}[h]
\plottwo{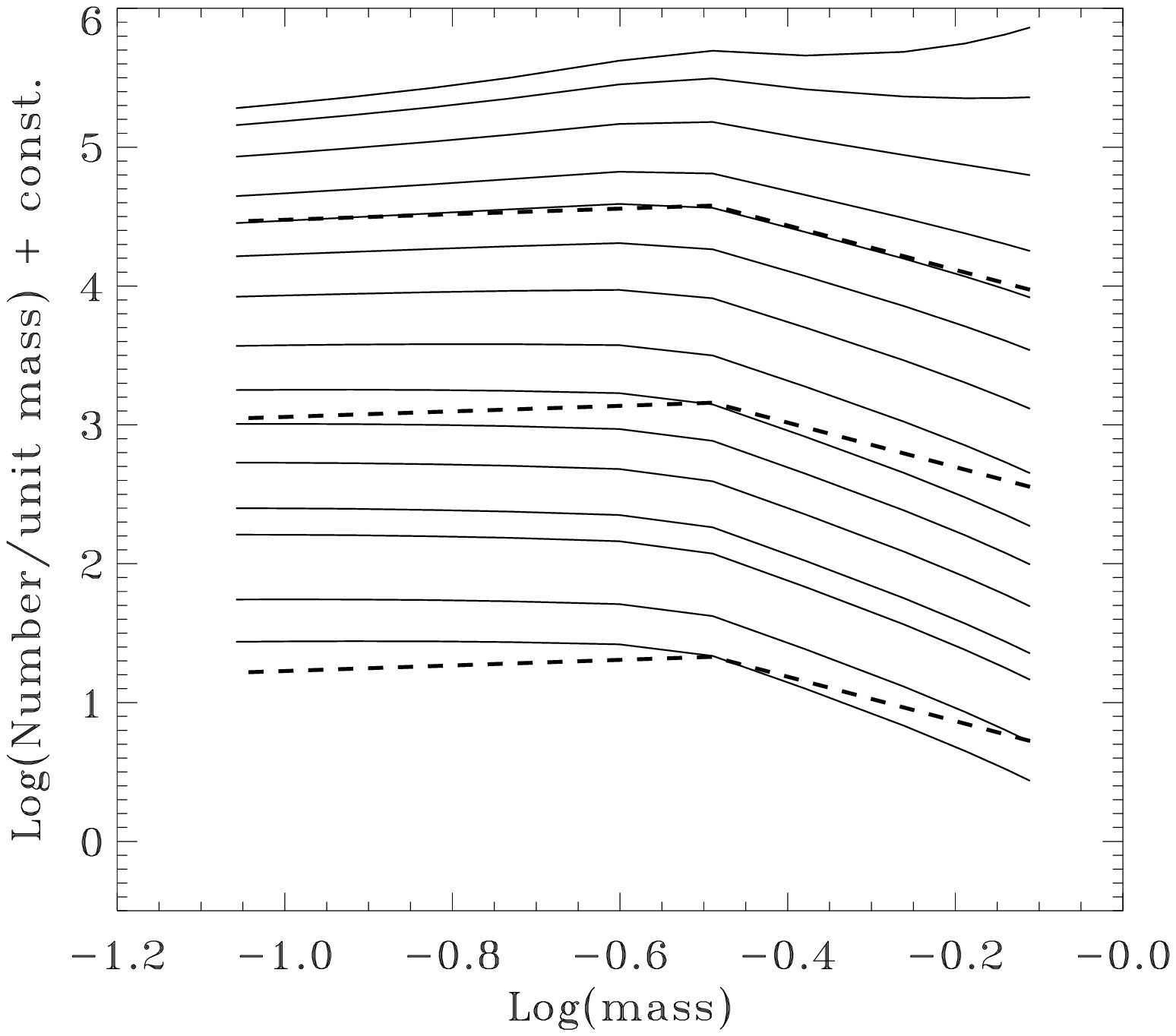}{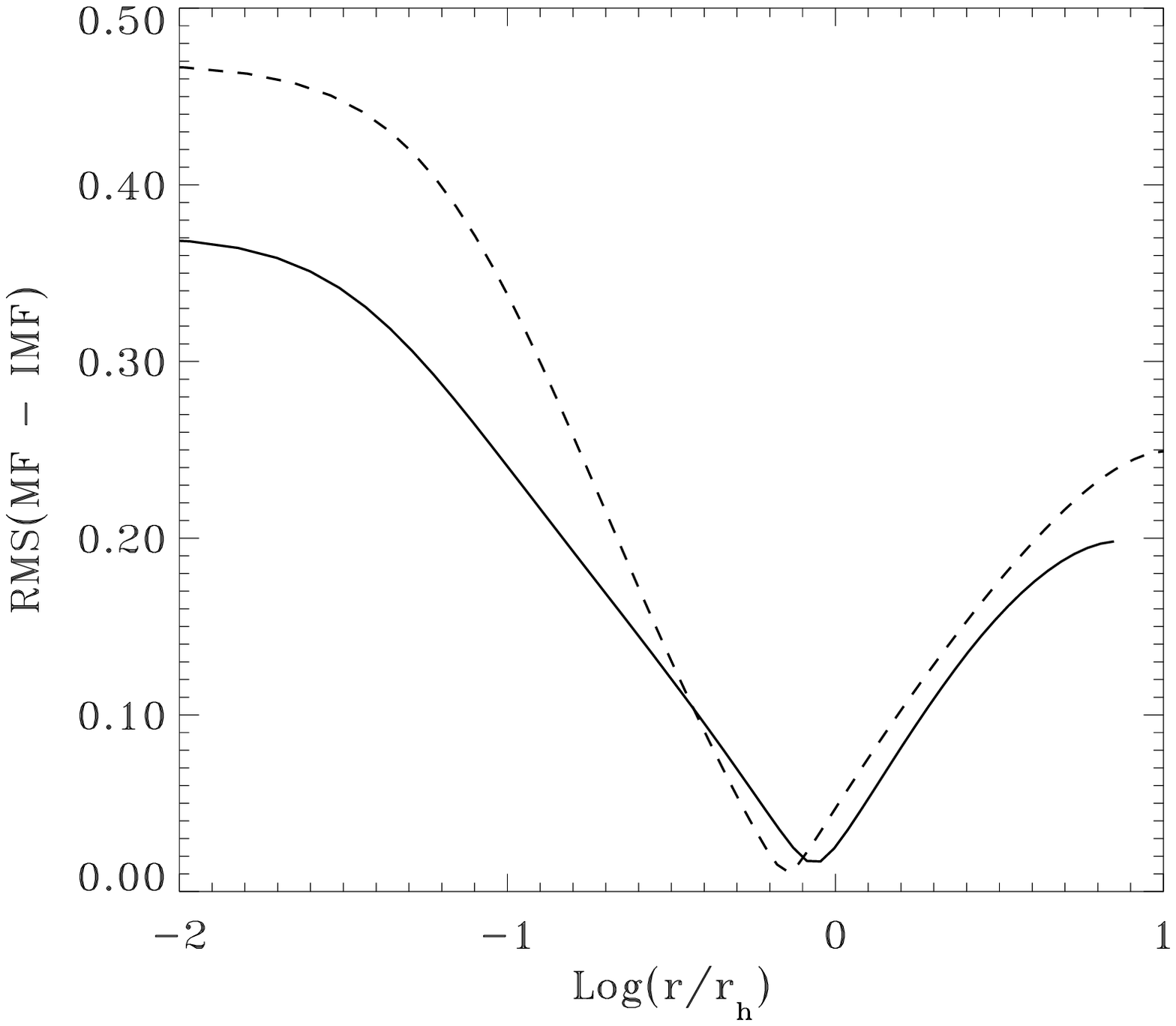}
\caption{Effects of mass segregation on the shape of the MF as
predicted by Michie--King models. {\it Left panel:} The expected local
MF is plotted as a function of radial position and compared to the
input global MF (dashed lines) for NGC\,6397. Radial distances are, from
top to bottom,  $0, 0.1, 0.3, 0.6, 1, 1.5, 2, 3, 4, 5, 6, 7, 8, 9, 10$
times $r_h$. {\it Right panel:} standard deviation of the
local from the global MF as a function of radial position for NGC\,6397
(solid line) and NGC\,104 (dashed line).}
\end{figure}

The MF shown in Figure\,3 need to be corrected for the effects of mass
segregation due to energy equipartition as has been extensively
discussed in Pulone et al. (1999), for example, in the case of
NGC\,6121 (M\,4).  To assess the magnitude of the expected effect on
the shape of the local MF, we have run standard multi-mass Michie--King
models for the clusters in our sample. As a typical example, in
Figure\,4 (left panel), the expected local MF is plotted as a function
of radial position and compared to the input global MF (dashed lines)
for the case of NGC\,6397. As can be seen in this figure, the largest
departure of the local MF from the global MF occurs in the innermost
regions of the cluster ($r \lesssim 0.1 r_h$) but significant
deviations also occur beyond the half-light radius. Near the half-light
radius the deviations are insignificant.This is shown graphically in
the right panel of Figure\,4, where we plot the standard deviation of
the local from the global MF as a function of radial position for this
cluster. These results are quite consistent with the expectations of
previous work along these lines by Richer et al.  (1991) and Pryor,
Smith, \& McClure (1986).

It is clear from this result that, provided the measurements are
carried out close to the half-light radius where the effects of mass
segregation are minimal, the deviation between the local MF and the
global one is basically unmeasurable with present techniques.  This
result was confirmed for NGC 6397 where the global MF was found to be
essentially indistinguishable from the MF determined at the half light
radius (De Marchi et al. 2000). While the majority of the LF in our
sample fulfill this requirement and can, therefore, be left safely
unaltered, those of NGC\,6341, NGC\,7078, and NGC\,7099 have been
obtained farther out in the cluster, at about $\sim 4\,r_h$ (see
Table\,1) in a region where the deviations can be significant as
indicated in Figure\,4. The effects of mass segregation must be
accounted for in these cases as they might otherwise lead to global MF
that appear steeper than they really are. The corrected $\Delta\,\log\,N$
for these clusters obtained by using the appropriate Michie-King models
are listed in Table\,2. The effect of this correction is a decrease, as
expected, of the value of $\Delta\,\log\,N$ for all three clusters.

\section{Physical Data and Tidal Disruption}

In Table\,2, we list the main physical parameters of the clusters
surveyed so far. Since our main objective is to search for a signature
of the cluster's dynamical history on its low mass MF, we have included
in this table whatever is known about its orbit in the Galactic tidal
field. The space motion data were obtained from the work of Dauphole et
al. (1996), Odenkirchen et al. (1997), and Dinescu, Girard, \& Van
Altena (1999). These data can be used in a theoretical model to
determine the change with time of the cluster's main characteristics
such as total mass, mass and luminosity functions, tidal radii, central
concentrations, relaxation times, etc. Both N-body and Fokker-Planck
models of increasing sophistication have been used recently to compute
such evolution (Takahashi \& Portegies Zwart 1999; Gnedin, Hernquist,
\& Ostriker 1999; Vesperini 1998, 1997; Vesperini \& Heggie 1997;
Gnedin \& Ostriker 1997; Capriotti \& Hawley 1996; Murali \& Weinberg
1997). Although different authors use different initial conditions and
approximations to the complex tidal interaction mechanisms, the
generally physically plausible final result is a flattening of an
assumed power-law low mass MF with time due to the preferential
evaporation of lower mass stars forced by two-body relaxation out to
the cluster periphery where the evaporation process is accelerated by
tidal shocks.

A direct calculation of this phenomenon for a specific cluster orbit
has not been carried out yet but an indirect indication at least of the
magnitude of the effect can be gleaned from the recent calculations of
the time to disruption $T_d$ of specific clusters carried out by Gnedin
\& Ostriker (1997) and by Dinescu et al. (1999). These
times are given in Gyr in Table\,2 (assuming a value of 10 Gyr for a
Hubble time) where the two values of the total destruction rate given
by Gnedin \& Ostriker (1997) for the two galactic models used in their
calculations have been averaged in column (10). The observed clusters
cover quite a large range of $T_d$ from a minimum of 4\,Gyr for
NGC\,6397 to 213\,Gyr for NGC\,5272 (using Gnedin \& Ostriker's
values), or from 2\,Gyr for NGC\,6121 to 275\,Gyr for NGC\,5272
(following Dinescu et al. 1999). These values should in principle be
regarded as upper limits to the true $T_d$, as both Gnedin \& Ostriker
and Dinescu et al. treat the internal dynamical evolution of the
clusters by using single-mass Michie--King models and, thus, tend to
underestimate the effects of mass segregation.  Although differences
exist between the values of $T_d$ as given by Gnedin \& Ostriker and
Dinescu et al. (with the latter being usually larger), an inspection of
Table\,2 shows that no one particular orbital parameter or the cluster
mass by itself is sufficient to foretell what the fate of the cluster
will be.  Even, for example, the cluster's perigalactic distance or its
height above the plane are not well correlated with $T_d$. This means
that the overall impact of the repeated bulge and disk shocks on the
cluster over its lifetime is not easily predictable from a simple
glance at the orbital parameters but only from the use of calculations
over the entire orbit such as those referred to above.

\begin{figure}[h]
\plotone{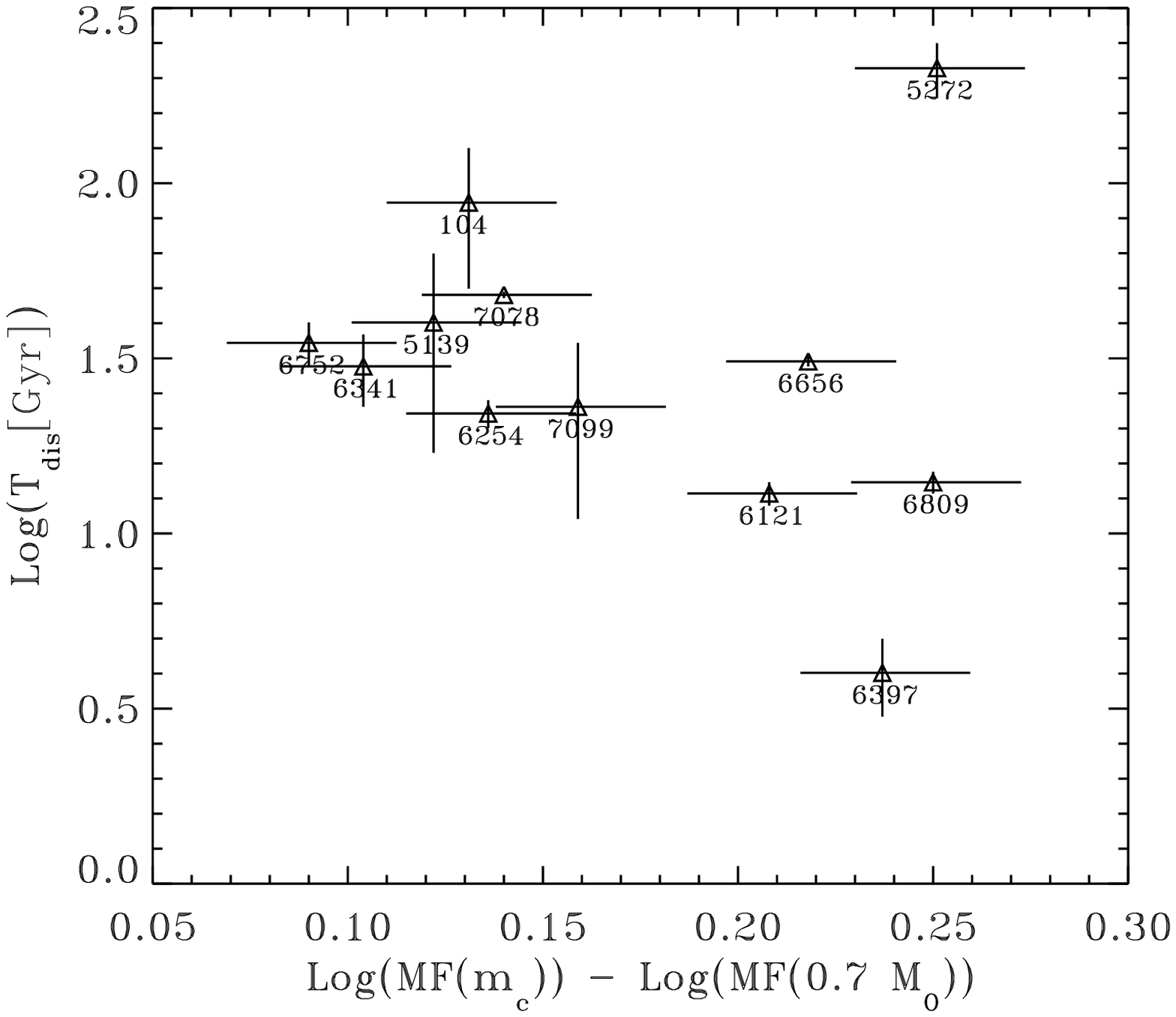}
\caption{The time to disruption in Gyr as calculated by Gnedin \&
Ostriker (1997) is shown here as a function of the index $\Delta\,\log\,N$
describing the shape of the MF in Figure\,3 (see text). Vertical error
bars reflect the difference between the values of $T_d$ obtained with
two different models describing the distribution of stars in the
Galaxy.  Horizontal error bars mark the range in which $\Delta\,\log\,N$ can
vary as a result of the uncertainty affecting the values of $m_c$ and
$\sigma$ in the MF. As such, both error bars define in practice a
$\pm 3\,\sigma$ uncertainty.}
\end{figure}

All things being equal, then, we would expect that the clusters with
the largest times to destruction $T_d$, i.e. those that have suffered the
least tidal disruption, to have the largest low to high mass number
ratio $\Delta\,\log\,N$. The actual situation is shown in Figure\,5 where we
have plotted the time to disruption of Gnedin \& Ostriker (1997) as a
function of $\Delta\,\log\,N$. The best linear fit to this distribution is a
straight line with zero abscissa at $T_d=113 \pm 3$ and having a slope
of $-2.9$ with a formal error of $\pm 0.2$. A horizontal line drawn at
$\log\,T_d \simeq 1.5$, however, would still give an acceptable fit.
Within the errors, then, there is no discernible trend in this
direction and the conclusion at this point is, therefore, quite clear:
the global MF of the clusters in our sample show no evidence of 
evolution with time within the quoted errors.

The MF logarithmic ratios $\Delta\,\log\,N$ plotted in Figure\,5 do seem to
have a statistically significant scatter about the mean beyond that
expected from measurement errors alone. What this scatter is due to is
not at all clear at the moment since we cannot simply ascribe it to
tidal effects. Intrinsic variations of the observed magnitude in the
IMF slope in this mass range are generated naturally in Adams \&
Fatuzzo's (1996) hypothesis due to variations in conditions under which
accretion is choked off by the appearance of winds from the
proto-star.  The same effect is predicted by Elmegreen's (1999a, 1999b)
hierarchical cloud sampling model due to both the inherently random
sampling process and the variation of the thermal Jeans mass with
initial conditions. If confirmed with more precise data, this effect,
if real, may be a sensitive indicator of natal cloud conditions such as
temperature and pressure.

In light of the very small range spanned by $m_c$ and $\sigma$ deduced
for all our clusters, the conclusion is, then, that a single form of
the MF can easily reproduce all the 12 deep LF obtained so far and,
since there is no obvious dependence on dynamical history over an
extremely wide range of conditions, that this MF is most likely to
represent the initial distribution of stellar masses in the cluster,
namely the IMF.

Finally, we should note that, although we have argued on the basis of
the data presented above that there seems not to be any, as yet,
measurable effect of tidal interactions with the galaxy showing up at
or just beyond the half-light radius and implying a massive ongoing
disruption event in any of the 12 clusters in our sample, there is
evidence of this effect in the LF of NGC\,6712 as recently measured
with the VLT (De~Marchi et al. 1999). This cluster's MF, if
extrapolated to the relevant mass range of the others, would show a
$\Delta\,\log\,N \simeq -0.1$. This result implies that some clusters are
much more capable than others in shielding  their interiors very
effectively from tidal disruption while others, like NGC 6712 are very
vulnerable to this effect.

How this may work in practice is starting to be understood by recent
theoretical and numerical simulation studies (Takahashi \& Portegies
Zwart 1999; Gnedin, Hernquist \& Ostriker 1999). For example, such
calculations do predict that most of the clusters in our present sample
are quite stable being located well inside the survival boundaries of
the vital diagrams plotted by Gnedin \& Ostriker (1997). They may have
survived so far relatively undisturbed in the interior at least due to
special initial conditions (high mass and concentration) and a
relatively benign shock history even though their outer parts may well
show indications of tidal heating (Drukier et al. 1998; Leon, Meylan \&
Combes 1999). In any case, they are unlikely to have lost more than
$\sim 1\,\%$ of their mass due to tidal shocking (Combes, Leon \&
Meylan 1999). NGC\,6712, on the other hand, may be one of the few
caught in the brief period before complete disruption. Takahashi \&
Portegies Zwart (1999) predict that, initially, NGC\,6712 had more than
1000 times its present mass.

\section{Discussion}

\subsection{Comparison with Previous Work}

A preliminary comparison of the properties of the first deep cluster LF
measured with the HST was carried out by De Marchi \& Paresce (1997,
and references therein). In those papers, we showed that the shape of
the LF, all measured near the half-light radius, seemed to bear little
or no relation with the past dynamical history of the clusters nor with
their position in the Galaxy. In spite of the widely different
dynamical properties of the low-metallicity clusters NGC\,6397,
NGC\,6656, and NGC\,7078, near their half-light radius these three
objects feature what in practice is the same LF below $\sim
0.6$\,\Msolar (see Figure\,1). Our conclusions, then, on the basis of a
much more limited data base and uncertain models were fortuitously
substantially similar to those in this paper.

From their comparison of the LF of nine clusters (all of which are
also part of the present study), Chabrier \& M\'era (1997) concluded
that the MF of GC is ``well described by a slowly rising power-law
$dN/dm \propto m^{-\alpha}$ with $0.5 \lesssim \alpha \lesssim 1.5$
down to $0.1$\,\Msolar,'' at variance with what we show in Figure\,3.
Several factors might be at the origin of this discrepancy. First, in
order to compare LF measured in different wavelength bands, Chabrier \&
M\'era converted them all into bolometric luminosities.  Although the
claim is that the LF of the same cluster observed through different
filters should yield the same bolometric LF, mixing data with theory
makes the true uncertainty much more difficult to estimate.  Second,
two of the LF that they used are now known to be incorrect at the low
mass end, namely that of NGC\,6397 of Cool et al. (1996) later amended
in King et al. (1998) and that of NGC\,5139 measured by Elson et al.
(1995) and recently corrected by De Marchi (1999). Both LF
overestimated the number of objects below $\sim 0.3$\,\Msolar, thus
mimicking an increase in the MF where a flattening should have occurred
instead.

A third effect, partly ensuing from the second, is that, having noticed
the discrepancy existing then between the LF of NGC\,6397 of Cool et
al. and that of Paresce et al. (1995) at the low mass end, Chabrier \&
M\'era were forced to exclude NGC\,6397 from their analysis. But since
the LF of NGC\,6397 reaches the faintest luminosities observed so far,
ignoring it prevents a reliable determination of the MF at the lowest
mass end.  Finally, they also ignore the flattening of the MF of
NGC\,6656 and NGC\,6341 below $0.2$\,\Msolar. All of this, combined
with the considerable structure that one sees in the MF above $\sim
0.3$\,\Msolar, makes any claim based on a single exponent power-law
mass distribution extending all the way to $\sim 0.1$\,\Msolar for GC
presently unsustainable. The only way to modify this conclusion would
be to assume an error in the Baraffe et al. (1997) and Cassisi
(1999) ML relations at the lowest masses. The only known reason for a
discrepancy would be in the possible formation of dust in the
atmosphere but this effect should be minimal in stars of such a low
metallicity.

Piotto et al. (1997) have also noticed the unsuitability of a single
power-law distribution to represent the MF of GC. In fact, the MF that
they obtain by applying the ML relation of D'Antona \& Mazzitelli
(1995) or that of Alexander et al. (1997) to their LF deviate from a
single exponent power-law, even when they restrict their investigation
to the small mass range below $\sim 0.4$\,\Msolar. As Figure\,5 and
Table\,2 clearly show, this is not unexpected because the peak of the
MF is located at $m_c\simeq 0.33$\,\Msolar. A drop-off below the peak 
at $\sim 0.3$\,\Msolar is also found by King et al. (1998) whose LF, 
if anything, is slightly steeper than ours (see the comparison shown in
Figure\,5 of De Marchi et al. 2000).

Piotto \& Zoccali (1999) again use a power-law fit to the MF of a
larger sample of clusters --- even if strong departures from a pure
power-law distribution are clearly evident at both the higher and lower
mass ends --- to claim the existence of a correlation between the rate
of destruction of their sample of clusters with the slope of their best
fit power law to the MF. On the other hand, restricting, for example,
Figure\,5 only to the clusters studied by Piotto \& Zoccali (1999)
would not show any correlation between the time to disruption of a
cluster with $\Delta\,\log\,N$. The reason for this discrepancy is not clear
but it may have to do with a combination of smaller sample, a power law
slope that bears little relation with the mass distribution of the
stars in the clusters, and rough estimates of the effects of mass
segregation. This last point is absolutely essential for proper
inter-cluster comparisons especially in view of the fact that most
HST--WFPC\,2 LF tend to be taken at constant sky offsets ($\sim
4\farcm5$) from the core and that, therefore, the farther away the
cluster the farther out physically is the LF sampled and the
segregation correction greatest.

Another problem commonly encountered in this type of work is that what
seem like significant differences in the LF are almost completely
washed out in the MF from which they descend especially once the
effects of mass segregation are accounted for. Thus, no meaningful
comparison between clusters can be made on the basis of the LF alone
and, since possible effects of evaporation are impressed on the MF, it
is exclusively in this plane that they must be sought.

Finally, Silvestri et al. (1998) argue that the MF of the entire MS of
NGC\,6397 is consistent with a shallow slope power-law if their most
recent models for the ML relation are used to convert the cluster LF
shown in Figure\,1. As these authors point out, this is due to their ML
relation being steeper than that of Baraffe et al. (1997) at the low
end of the MS. As we already mentioned above, these models rely on a
grey atmosphere approximation that is not self-consistent and that,
therefore, cannot be preferred to the latter that rely on a completely
self-consistent approach which provides excellent matches to a wide
variety of experimental data.

Silvestri et al. (1999) also address the issue of how a change in the
distance scale of globular clusters will affect their MF, showing that
longer distances result in shallower MF, or, better, in a more
pronounced flattening at the low-mass end. In view of the still debated
revision of the distances to globular clusters based on the new
Hipparcos data (see e.g. Gratton et al. 1997; Pont et al. 1997), we
have adopted the pre-Hipparcos values in our analysis. Nevertheless,
since the distance moduli of the 9 globular clusters studied by Gratton
et al. (1997) are in excess of those of Djorgovski (1993) by only $0.22
\pm 0.10$\,mag, using the new scale would not change our results
significantly.

\subsection{Comparison with Other Clusters}

\begin{figure}[h]
\plotone{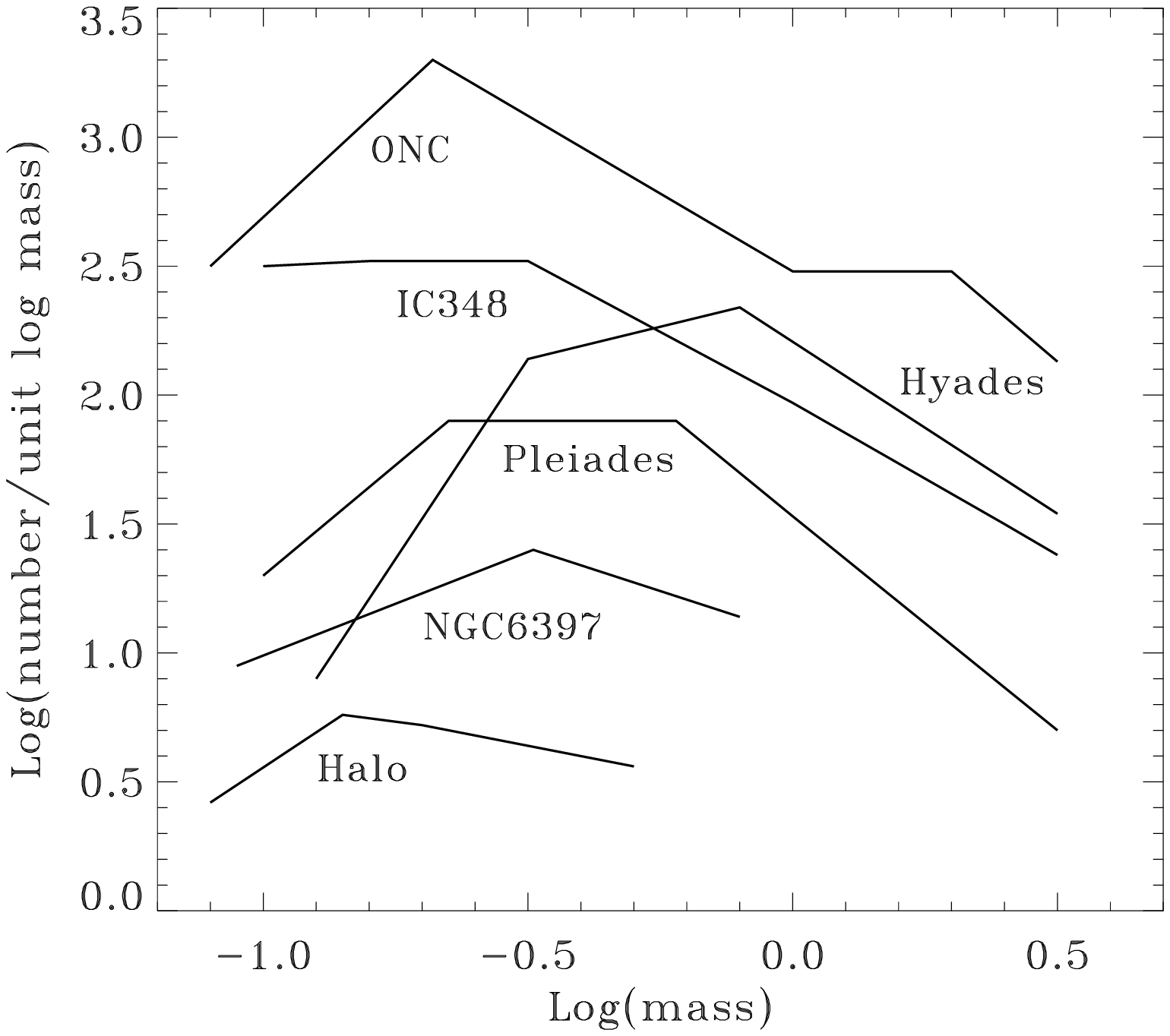}
\caption{Mass function of stars in young open clusters, in the Halo,
and in NGC\,6397 (see text).}
\end{figure}

How does the globular cluster IMF derived here compare with MF derived
in other stellar clusters?  The present situation in this regard is
summarized in Figure\,6 where we have drawn the most recently deduced
MF in several Galactic cluster populations. The data are for the Orion
Nebula Cluster from Hillenbrand (1997), IC348 from Luhman et al. (1998)
and Herbig (1998), the Pleiades as reported by Bouvier et al. (1998),
and the Hyades from Reid \& Hawley (1999) and Gizis, Reid \& Monet
(1999). For comparison, the MF deduced from the NGC\,6397 LF shown in
Figure\,3 is also reproduced here in the same format. Since the
Galactic halo may well be populated mainly by the disruption of
globular clusters, we also show in this figure the MF of the halo as
obtained from the LF of Dahn et al. (1995) confirmed by Fuchs \&
Jahreiss (1998) and Gizis \& Reid (1999), converted to a MF by Chabrier
\& M\'era (1997), and corrected for binaries by Graff \& Freese
(1996).

The measurement peculiarities and sources of uncertainties in these
measurements are exhaustively discussed and quantitatively evaluated by
these authors so that they should be regarded as the most precise and
up to date determinations of the MF in the $\sim 0.1 - 3$\,\Msolar mass
range.  We have distilled their measurements into the best
fitting power law in the appropriate mass range. The data are shown in
logarithmic mass units on the abscissa and the MF represented as the
$\log dN/d\log m$ in the ordinate. The slope of the power law is then
$-x=1-\alpha$ where $\alpha$ is the slope in linear mass units
($\alpha=2.35$ for the Salpeter IMF). The vertical position of the
lines is arbitrary, of course, so we have shifted them up or down for
enhanced visibility.

Errors associated with the measurements of the slopes and masses of
several tenths of a dex in both axes should be considered typical. It
should also be noticed that, in many cases, the number of objects in
the faintest bins is very small ($\sim 1 - 2$). The effect of
unresolved binaries on the MF is estimated by most authors but it does
remain as a caveat to keep in mind until much higher spatial resolution
observations become available (Kroupa 1995). Mass segregation may also
play a role in some cases like the Hyades which may explain the higher
than average $m_c$ for this cluster.
 
A striking aspect of the results shown in Figure\,6 is the similarity
between the various MF despite the substantial differences in
environment and physical characteristics such as metallicity and age.
To be sure there are possible variations and inconsistencies in the
details but overall the trend is pretty clear, namely a Salpeter-like
increase in numbers with decreasing mass from $\sim 3$\,\Msolar to
$1$\,\Msolar always followed by a definite break and flattening
extending down to $0.1 - 0.2$\,\Msolar with slopes in the range $0 < x
< 0.5$. Moreover, all the measurements that reach close to the
H-burning limit with reasonable completeness and statistical
significance indicate a turnover below $\sim 0.2$\,\Msolar. Certainly
no one simple power-law can possibly explain the data shown in this
figure, thus ruling out a scale free IMF in any of these cases.

It is quite conceivable, then, that, at least to the level of accuracy
of the present data, all the MF schematically represented in this
figure  come from basically the same underlying type of distribution
function that increases with mass from the substellar limit to a peak
somewhere between $0.2$ and $0.5$ solar masses and then drops steeply
beyond $\sim 1$\,\Msolar.  More specifically, convolving a log-normal
distribution function with the limited mass resolution and counting
statistics presently available could easily generate the segmented
power law MF shown in Figure\,6.  In other words, there seems to be no
reason to think that the shape of the IMF from which the various
samples are taken is much different from a log-normal implied by the
globular cluster data. It is possible that all the MF of the clusters
shown in Figure\,6 could be the result of a single log-normal since the
evident cluster to cluster scatter of peak mass could be completely due
to measurement error or systematic effects like mass segregation.  In
this case, the log-normal implied by the globular cluster data would be
a truly universal function essentially independent of age or
metallicity.

Another possibility is that this scatter is real and related to some
fundamental characteristic of the cluster. The small number of clusters
for which reliable MF have been obtained so far precludes, for the
moment, precise conclusions such as whether or not there is a trend for
$m_c$ in clusters to increase with age as one might be attempted to
deduce from the data in Figure\,6. Moreover, there is a group of
embedded clusters like $\rho$\,Oph and NGC\,2024 that do show a
steadily rising MF all the way from $\sim 1$\,\Msolar to well below the
H-burning limit at $\sim 0.08$\,\Msolar (Williams et al. 1995; Comeron
et al. 1996). These MF taken at face value cannot be reconciled with
the log-normal form discussed so far unless their $m_c$ is located deep
in the lower reaches of the brown dwarf regime.  The Pleiades
themselves seem to have a rising MF with decreasing mass below the
H-burning limit even if the stellar part is well represented by a
log-normal distribution (Bouvier et al. 1998).

These cases would argue for a non universal IMF which could be
sensitive to peculiar physical conditions affecting the formation of
very low mass stars and brown dwarfs (Evans 1999; Liebert 1999). It is
possible to entertain the idea, for example, that physical conditions
in dense, massive clusters like the ONC or the globular clusters
discussed here are not conducive to their formation. Because there are
many open issues surrounding the accurate determination of MF in
embedded clusters by means of LF modelling (Luhman et al. 1998; Meyer
et al. 1999), and because it is still difficult to pin down the mass
peak in young clusters with great enough accuracy, it is probably still
too early to tell if this is a serious hypothesis or not but it does
raise at least the very exciting possibility of using the bottom of the
MS as a sensitive diagnostic of initial conditions in the original star
forming regions.

\subsection{Comparison with Theory}

 How plausible is a log-normal distribution of the kind advocated here
from a purely theoretical perspective? As first pointed out by Larson
(1973) and Zinnecker (1984), when the star formation process depends on
a large number of independent physical variables, the resulting IMF can
be approximately described by a log-normal distribution function of the
form discussed in the previous section. As developed in greater detail
recently by Adams \& Fatuzzo (1996), the observed values of the mass
scale $m_c$ and the standard deviation $\sigma$ can even be used to set
rough limits on the actual physical variables entering into the theory
if, as they claim, the mass of a star is self-determined by the action
of an outflow. The values of these two parameters obtained for the
globular cluster sample discussed in the previous section are quite
consistent with our present, admittedly limited, knowledge of the
conditions in the star-forming environment. In general, in this
particular formulation of the theory, very low mass stars and brown
dwarfs are relatively rare since they require natal clouds having
unrealistically low effective sound speeds.

On the other hand, it is well known that the IMF cannot be {\it
completely} described by a log-normal form since it is very unlikely
that so many variables are involved in the formation process and the
greatest deviations will be in tails at the extremes of the function.
Thus, it is still quite plausible theoretically to have cases where the
lowest mass end of the IMF deviates even significantly from the
log-normal form as in the case of the embedded clusters NGC\,2024 and
$\rho$\,Oph discussed above. Hierarchical fragmentation may be quite
relevant in setting the form of the IMF in the low mass range discussed
in this paper (Larson 1995) and this process also would be expected to
yield, in principle, a log-normal IMF under the proper circumstances.
Recent numerical simulations of the formation of proto-stellar cores
from the collapse of dense, unstable gas clumps and subsequent
evolution through competitive accretion and interactions such as those
expected in a dense cluster, predict a mass spectrum described by a
log-normal function quite similar to the ones derived in Figure\,6
(Klessen \& Burkert 1999) lending even more support to the idea that
these represent the original mass function of these clusters.

A completely different purely mathematical approach taken by Elmegreen
(1997, 1999a, 1999b) recently arrives at very similar conclusions as to
the form of the underlying stellar IMF. In this formalism, proto-stellar
gas is randomly sampled from clouds with self-similar hierarchical
structure giving rise to an IMF that looks remarquably similar to the
one outlined in the previous section. This includes a power law section
at intermediate masses and a flattening and turn-over at low masses due
to the inability of gas to form stars below the thermal Jeans mass. Of
particular interest in our context here is the natural occurrence of
IMF fluctuations of several tenths slope due to random variations
around a universal IMF quite similar to those observed for the
$\Delta\,\log\,N$ of our cluster sample. This theory would then quite
naturally explain the scatter observed in this parameter shown in
Figure\,5. As more data is gathered, if this finding is confirmed it
could be used as a strong constraint on theory. On the other hand, such
a scatter acts to obscure or even obliterate any sign of possible tidal
effects on the MF and again explains why these effects are not at all
evident in the data shown in Figure\,5.  Since the thermal Jeans mass
does depend somewhat on environmental conditions, this theory might
also be able to explain the possible inter-cluster variation of $m_c$
seen in Figures\,3 and 6.

\section{Summary and Conclusions}

We have analyzed in detail the implications for the IMF of our present
reasonably good knowledge of the MS LF of a dozen Galactic globular
clusters covering a wide range of physical and orbital
characteristics.  We have shown, first, that they can be converted to a
MF by the application of a ML relation that incorporates all
the relevant internal and atmospheric physics of low mass low
metallicity stars appropriate to the cluster sample under
investigation. We have, then, calculated the possible effect of mass
segregation due to energy equipartition on the locally derived MF and
find that for nine of the twelve clusters no correction is required as
they were obtained very close to the half-light radius where the
deviation is negligible. For the other three clusters, corrections are
applied that reduce the observed number gradient between $0.7$\,\Msolar
and the mass peak.

The MF obtained in this way could all descend from the same log-normal
form of the global MF within a small range of mass scales and standard
deviation. The MF of the four clusters of the sample whose LF extend
significantly beyond the mass peak at $0.33$\,\Msolar cannot be
reproduced by a single power law throughout the MS mass range explored
in this paper, but would require at least an unphysical double power
law.  We, then, explored the possible modification of these global MF
with orbital history of the individual clusters by comparing the number
gradient of their MF with theoretical estimates of their survivability
in the Galactic potential. No statistically significant effect is
found, no matter what particular model is used. We conclude that the
effect, if present at all in this type of clusters, is completely
obscured by the present observational uncertainties and that,
therefore, the global MF we measure today must be, within those
uncertainties, identical to the original MF namely the IMF.

We explored, finally, the plausibility of this conclusion by examining
the measured structure of the MF of much younger clusters that could
shed light on the shape of the original globular cluster MF. For many
of the best measured clusters, we find convincing evidence that they
also exhibit a log-normal shaped IMF in the stellar mass range of the
same type deduced from the globulars, albeit, possibly, with a wider
range of mass scales and standard deviations. Both the shape and the
scatter are roughly consistent with presently available theoretical
models.  A few deeply embedded clusters do show evidence of possible
deviations from this result although there are still some questions as
to the validity of the measurement techniques in these difficult
environments.

Thus, the conclusion seems robust at this point that most cluster stars
originate from a quasi-universal IMF below $1$\,\Msolar having the
shape of a log-normal whose precise mass scale and standard deviation
may fluctuate from one particular environment to another due to the
effects of random sampling or differing physical conditions depending
on which model is appropriate. It is also clear that much remains to be
done to clarify and establish the range of validity of this conclusion
and to understand the origin of the possible deviations such as those
found for some embedded clusters. This investigation should yield a
bountiful harvest of information on the stellar IMF in the near
future.  Of particular importance in this endeavour, will be securing a
sufficiently large, clean sample of stars of the same physical and
kinematical type in a wide variety of environments and ages and to
develop the most accurate models of their energy output as a function
of mass.
 
\acknowledgments

We would like to thank France Allard, Isabelle Baraffe, Santi Cassisi,
Gilles Chabrier, Dana Dinescu, Bruce Elmegreen, Oleg Gnedin, Pavel
Kroupa, and Simon Portegies Zwart for useful discussions and an
anonymous referee for comments and suggestions that significantly
strengthened the paper. We are particularly grateful to Luigi Pulone
for having computed the effects of mass segregation on the MF of the
clusters in our sample.


\begin{references}

\reference{} 
Adams, F.C., and Fatuzzo, M. 1996, ApJ, 464, 256

\reference{} 
Alexander, D., Brocato, E., Cassisi, S., Castellani, V., Ciacio, F.,
and Degl'Innocenti, S. 1997, A\&A, 317, 90

\reference{} 
Baraffe, I., Chabrier, G., Allard, F., and Hauschildt, P. 1997, A\&A,
327, 1054

\reference{}
Baraffe, I., Chabrier, G., Allard, F., and Hauschildt, P. 1998, A\&A,
337, 403

\reference{}
Bouvier, J., Stauffer, J.R., Martin, E.L., Barrado y Navascues, D.,
Wallace, B., and Bejar, V.J.S. 1998, A\&A, 336, 490

\reference{}
Capriotti, E.R., and Hawley, S.L. 1996, ApJ, 464, 765

\reference{}
Carretta, E., et al. 1999, in preparation

\reference{}
Cassisi, S. 1999, private communication

\reference{}
Chabrier, G., Baraffe, I., Allard, F., and Hauschildt, P. 1999, in From
Extrasolar Planets to Brown Dwarfs, ESO Conf. Proc., VLT Opening
Symposium, ed. F. Paresce, astro-ph/9905210

\reference{}
Chabrier, G., Baraffe, I., and Plez, B. 1996, ApJ, 459, 91

\reference{}
Chabrier, G., and M\'era, D. 1997, A\&A, 328, 83

\reference{}
Combes, F, Leon, S., and Meylan, G. 1999, A\&A, in press

\reference{}
Comeron, F., Rieke, G.H., and Rieke, M.J. 1996, ApJ, 473, 294

\reference{}
Cool, A. M., Piotto, G., and King, I. R. 1996, ApJ, 468, 655 

\reference{}
Dahn, C., Liebert, J., Harris, H.C., and Guetter, H.H. 1995, in 
The bottom of the Main Sequence and Beyond, ESO Conf. Proc., ed. 
C.G. Tinney (Berlin: Springer), 239

\reference{}
D'Antona, F., and Mazzitelli, I. 1996, ApJ, 456, 329

\reference{}
Dauphole, B., Geffert, M., Colin, J., Ducourant, C., Odenkirchen, M.,
and Tucholke, H.J. 1996, A\&A, 313, 119

\reference{}
De Marchi, G. 1999, AJ, 117, 303

\reference{}
De Marchi, G., and Paresce, F. 1995a, A\&A, 304, 202 

\reference{}
De Marchi, G., and Paresce, F. 1995b, A\&A, 304, 211 

\reference{}
De Marchi, G., and Paresce, F. 1996, in Science with the Hubble Space
Telescope - II, Eds. P. Benvenuti, F.D. Macchetto, \& E. Schreier
(Baltimore:  STScI), 310

\reference{}
De Marchi, G., and Paresce, F. 1997, ApJ, 476, L19 

\reference{}
De Marchi, G., Paresce, F., and Pulone, L. 2000, ApJ, Feb 10 issue,
Astro-ph/9908251

\reference{}
De Marchi, G., Leibundgut, B., Paresce, F., and Pulone, L. 1999, A\&A, 343, L9

\reference{}
Dinescu, D.I., Girard, T.M., and Van Altena, W.F. 1999, AJ, 117, 1792

\reference{}
Djorgovski, S. 1993, in Structure and Dynamics of Globular Clusters, 
ASP Conf. Ser. 50, ed. S. Djorgovski and G. Meylan (San Francisco: ASP), 373

\reference{}
Drukier, G.A., Slavin, S.D., Cohn, H.N., Lugger, P.M., Barrington,
R.C., Murphy, B.W., and Seitzer, P.O 1998, AJ, 115, 708

\reference{}
Elmegreen, B.G. 1997, ApJ, 486, 944

\reference{}
Elmegreen, B.G. 1999a, in Unsolved Problems in Stellar Evolution, ed.
M. Livio (Cambridge: Cambridge University Press), in press,
astro-ph/9811289

\reference{}
Elmegreen, B.G. 1999b, ApJ, 515, 323

\reference{}
Elson, R.A.W., Gilmore, G.F., Santiago, B.X., Casertano, S. 1995, AJ,
110, 682

\reference{}
Evans, N.J. 1999, astro-ph/9905050

\reference{}
Ferraro, F.R., Carretta, E., Bragaglia, A., Renzini, A., and Ortolani,
S. 1997, MNRAS, 286, 1012

\reference{}
Fuchs, B., and Jahreiss, H. 1998, A\&A, 329, 81

\reference{}
Gizis, J.E., and Reid, I.N. 1999, AJ, 117, 508

\reference{}
Gizis, J.E., and Reid, I.N., and Monet, D.G. 1999, AJ, 118, 997

\reference{}
Gnedin, O.Y., Hernquist, L.,  and Ostriker, J. P. 1999, ApJ, 514, 109

\reference{}
Gnedin, O.Y., and Ostriker, J.P. 1997, ApJ, 474, 233

\reference{}
Graff, D., and Freese, K. 1996, ApJ, 456, L49

\reference{}
Gratton, R., Fusi Pecci, F., Carretta, E., Clementini, G., Corsi, C.,
and Lattanzi, M. 1997, ApJ, 491, 749

\reference{}
Hillenbrand, L.A. 1997, AJ, 113, 1733

\reference{}
Herbig, G. 1998, ApJ, 497, 736

\reference{}
King, I. R., Anderson, J. , Cool, A. M., and Piotto, G. 1998, ApJ, 492, 
L37 

\reference{}
Klessen, R.S., and Burkert, A. 1999, ApJ, in press, astro-ph/9904090

\reference{}
Kroupa, P. 1995, MNRAS, 277, 1522

\reference{}
Larson, R.B. 1973, MNRAS, 161, 133

\reference{}
Larson, R.B. 1995, MNRAS, 272, 213

\reference{}
Leon, S., Combes, F, and Meylan, G. 1999, A\&A, in press

\reference{}
Liebert, J., 1999, in Unsolved Problems in Stellar Evolution,  ed. M.
Livio (Cambridge: Cambridge University Press), in press,
astro-ph/9812061

\reference{}
Luhman, K., Rieke, G.H., Lada, C.J., and Lada, E.A.  1998, ApJ, 508,
347

\reference{}
Marconi, G., Buonanno, R., Carretta, E., Ferraro, F.R., Fusi Pecci, F.,
Montergriffo, P., De Marchi, G., Paresce, F., and Laget, M. 1997,
MNRAS, 293, 479

\reference{}
Meyer, M.R., Adams, F., Hillenbrand, L., Carpenter, J., and Larson, R. 
1999, in Protostars \& Planet IV, eds. V. Mannings, A. Boss, and S. 
Russell (Tucson: The University of Arizona Press), in press

\reference{} 
Meylan, G. 1987, A\&A, 184, 144

\reference{} 
Meylan, G. 1988, A\&A, 203, 297

\reference{}
Murali, C., and Weinberg, M.D. 1997, MNRAS, 291, 717

\reference{}
Odenkirchen, M., Brosche, P., Geffert, Tucholke, H.J. 1997, NewA, 2,
477

\reference{}
Paresce, F., De Marchi, G., and Romaniello, M. 1995, ApJ, 440, 216

\reference{}
Piotto, G. , Cool, A. M., and King, I. R. 1997, AJ, 113, 1345

\reference{}
Piotto, G., and Zoccali, M., 1999, A\&A, 345, 485

\reference{}
Pont, F., Mayor, M., Turon, C., and Vandenberg, D.A. 1998, A\&A, 329, 87

\reference{}
Pryor, C., Smith, G.H., and McClure, R.D. 1986, AJ, 92. 1358

\reference{}
Pulone, L., De Marchi, G., and Paresce, F. 1999, A\&A, 342, 440

\reference{}
Richer, H.B., Fahlman, G., Buonanno, R., Fusi Pecci, F., Searle, L.,
and Thompson, I. 1991, ApJ, 381, 147

\reference{}
Rood, R.T., Carretta, E., Paltrinieri, B., Ferraro, F.R. Fusi Pecci,
F., Dorman, B., Chieffi, A., Straniero, O., Buonanno, R. 1999, ApJ, 
523, 752

\reference{}
Rubenstein, E.P., and Bailyn, C.D, 1999, ApJ, 513, L33

\reference{}
Scalo, J.M. 1998, in The Stellar Initial Mass Function, ASP Conf. Ser.
142, eds. G. Gilmore and D. Howell (San Francisco: ASP), 201

\reference{}
Scalo, J.M. 1999, in The Birth of Galaxies, Blois, France,
astro-ph/9811341

\reference{}
Silvestri, F., Ventura, P., D'Antona, F., and Mazzitelli, I. 1998, ApJ,
509,192

\reference{}
Takahashi, K., and Portegies Zwart, S. 1999, ApJ, submitted,
astro-ph/9903366

\reference{}
Vesperini, E. 1998, MNRAS, 299, 1019

\reference{}
Vesperini, E. 1997, MNRAS, 287, 915

\reference{}
Vesperini, E., and Heggie, D.C. 1997, MNRAS, 289, 898

\reference{}
Webbink, R.F. 1985, IAU Symp, 113, 541 

\reference{}
Williams, D.M., Comeron, F., Rieke, G.H., and Rieke, M.J. 1995, ApJ,
454, 144

\reference{}
Zinnecker, H. 1984, MNRAS, 210, 43

\end{references}
\end{document}